# Recovery of Architecture Module Views using an Optimized Algorithm Based on Design Structure Matrices

Jemerson Figueiredo Damásio[1], Roberto Almeida Bittencourt[1, 2] and
Dalton Dario Serey Guerrero[1]
[1]Department of Systems and Computing
UFCG - Federal University of Campina Grande
Campina Grande, PB - Brazil
[2]UEFS - State University of Feira de Santana
Feira de Santana, BA – Brazil
{jemerson,dalton}@dsc.ufcg.edu.br, roberto@uefs.br

## Abstract

*Design structure matrices (DSMs) are useful to represent high-level system structure, modeling interactions between design entities. DSMs are used for many visualization and abstraction activities. In this work, we propose the use of an existing DSM clustering algorithm to recover software architecture module views. To make it suitable to this domain, optimization has proved necessary. It was achieved through performance analysis and parameter tuning on the original algorithm. Results show that DSM clustering can be an alternative to other clustering algorithms.*

## 1. Introduction

Software architecture recovery has been an important research topic since the early nineties. It can be used for a variety of purposes, such as improving software comprehension, documenting legacy systems, serving as a starting point for reengineering processes, identifying components for reuse, migrating systems to software product lines, co-evolving architecture and implementation, checking compliance between architecture, low-level design and implementation, analyzing legacy systems and achieving graceful software evolution [11, 18].

A large body of research on architecture recovery has been built for around fifteen years. A recent work surveys 146 papers which are either considered influential or propose a specific approach to the architecture recovery problem [18]. This survey classifies previous work in terms of inputs, outputs, techniques, processes and goals. Regarding the techniques, they are classified into quasi-manual, semi-automatic and quasi-automatic. In the first, the software architect manually recovers abstractions with the help of a tool. The second automates repetitive tasks, instructing the tool to recover abstractions. And the last infers architectural knowledge directly from software artifacts.

Quasi-automatic techniques are of special interest, since they promise faster recovery just from analyzing existing knowledge in software artifacts. They usually take advantage of other techniques like formal concept analysis, clustering and dominance in order to discover abstractions. For they use specific criteria, these techniques end up imposing software architectures that respect such criteria [1] (e.g.: if the heuristics of a clustering algorithm demand high cohesion and low coupling, the recovery process ends up producing highly-cohesive and low-coupled decompositions).

A design structure matrix (DSM) is a matrix representation of system entities and their interactions. It can be used as a means to better organize design activities and layouts, as an efficient design visualization technique and as a basis for system abstraction and mathematical analysis [4]. DSMs can also be used for system clustering [3].

In this work, we present a novel use for a clustering algorithm entitled design structure matrix clustering, or simply, DSMC, to the specific aim of clustering software design entities to recover architecture module views. To the best of our knowledge, this algorithm has not yet been used to this purpose. DSMC is a variation of an algorithm first published by Idicula [5]. We also show how we improved the algorithm, optimizing it for performance reasons, and tuning its parameters to adapt it to the architecture recovery domain. Finally, a case study with real open source systems was accomplished in order to compare DSMC results with other clustering algorithms. Our results indicate that DSMC is outperformed by k-means, but it also outperforms two other known clustering algorithms.

The remainder of this paper is organized as follows. Section 2 describes the original DSMC algorithm. Section 3 presents the research methodology we used to improve the algorithm. Section 4 shows results for parameter tuning and for the case study. Related work is discussed in section 5 and conclusions are derived in section 6.

## 2. The design structure matrix (DSM)

The design structure matrix (DSM) is a square matrix that represents design entities both as rows and columns. Dependencies between entities are shown in matrix cells. The cell in row *i* and column *j* represents the dependency from entity *i* to entity *j*. With this simple structure, it is possible to visualize designs, derive abstractions and perform mathematical analysis. Due to their simplicity and mathematical power, DSMs have been used for modeling and analyzing different domains, from system architecture, organization structures and its interactions, to processes and activities networks and their information flow [3].

DSMs were first applied to software by Sullivan et al. [22] to quantify the value of modularity in software. Later, it has gained widespread use in the software engineering community, with work ranging from software architecture [10, 19] to software evolution [14].

DSMs have also been applied for clustering, initially by Idicula [5]. He proposed an algorithm to create clusters from a large set of mutually interdependent tasks. This algorithm was the basis for a modified algorithm proposed by Fernandez [5], and later improved by Thebeau [23]. In their theses, they clustered teams from a company in order to ease information exchange in a product development effort.

### 2.1. The DSM clustering algorithm

We first present the original design structure matrix clustering algorithm (DSMC), as described by Fernandez [5]. In order to better understand the algorithm, some important concepts are first presented. The rationale behind these concepts in the context of team coordination and software architecture is also explained.

#### 2.1.1. Total coordination cost

This function tries to capture, in a mathematical formula, the following heuristics [5]:
1. It is more convenient to address interactions between modules formally in a system team, rather than ignoring them hoping that they will be addressed informally by the teams alone.
2. The time or cost to address an interaction is proportional to the frequency or importance of the interaction.
3. It is easier for teams to interact in smaller groups, rather than in large ones.
4. The difficulty of managing a system team and the effectiveness to address the interactions between member teams increases with the number of teams.
5. For individual teams, the cost of being a member of system teams increases with the number of system teams addressing an interaction.

Software architecture modules and system teams are similar, in terms of these heuristics. It is not unusual to have each team developing one software module. Even if this is not the case, information hiding and encapsulation by means of software modules attempt to decrease the information load to be dealt with by developers, by reducing module size and the number of dependencies between modules. Thus, we may say that modularization, information hiding, encapsulation, high cohesion and low coupling are well represented in the above heuristics.

In summary, the goals of the *Total Coordination Cost* (*TCC*) are: *i)* making explicit the interactions between system modules, *ii)* privileging relevant interactions, *iii)* creating relatively small clusters and *iv)* keeping the number of clusters relatively small.

Mathematically, *TCC* is given by the formula below, and is the sum of all *Intra Cluster Costs* (*ICC*) with all *Extra Cluster Costs* (*ECC*).

$$TCC = \sum ICC + \sum ECC$$

$$ICC = (DSM_{j,k} + DSM_{k,j}) \, mod\_size_i^{powcc}$$

$$ECC = (DSM_{j,k} + DSM_{k,j}) \, DSM\_size^{powcc}$$

where:

| | |
|---|---|
| *TCC* | = total coordination cost |
| *ICC* | = interaction cost within a module |
| *ECC* | = interaction cost between modules |
| $DSM_{j,k}$ | = interaction between modules *j* and *k* |
| *DSM_size* | = the DSM size |
| $mod\_size_i$ | = numb. of elements in module *i* |
| *powcc* | = penalizes modules size |

*Intra Cluster Cost* accounts for the interaction that happens between two design entities that belong to the same module. *Extra Cluster Cost* accounts for the interactions that happen between two design entities that belong to different modules. Interaction within a module contributes little to the total cost in comparison with interactions between modules. Furthermore, parameter *powcc* controls module size by penalizing solutions with large modules.

### 2.1.2. Bids

*TCC* is the objective function to be minimized. Entities are moved between clusters, changing the objective function value. In order to find a better clustering, the algorithm promotes an auction where each cluster bids a value for the moving entity – which is randomly chosen.

The *bid function*, or simply *bid*, is used to find the best destination module for a given entity in order to reduce *TCC*. This function is a measure of how strong the bidding module wants to interact with the entity under auction. It takes into account the interaction strength between the module entities and the entity under auction and the module size. It is calculated as follows:

$$bid(module_i) = \frac{sum\_interactions^{powdep}}{module\_size_i^{powbid}}$$

where:
*bid*(*module$_i$*) = module *i* bid for the entity
*I* = module index
*sum_interactions* = sum of the interactions between the entity under auction and the entities in module *i*
*Powdep* = emphasizes the interactions
*Powbid* = penalizes module size

Even the best bid may not guarantee that *TCC* will decrease its value, simply because the moving entity may already be in the best cluster. Hence, the move will only happen if an improvement is achieved in the objective function.

### 2.2. Algorithm overview

The steps of the original algorithm are explained below, with details shown in Figure 1.
1. Initially, every member is considered a module itself (line 2).
2. Repeat the following while results are still improving, i.e., while *TCC* changes are greater than a specified threshold (line 6).
   2.1. For a multiple of *DSM_size* iterations, repeat the basic entity moves (line 7).
      2.1.1. Move randomly chosen entity to the module that offers the best *bid* (lines 8 to 10).
      2.1.2. If the new *TCC* is higher than the previous one, undo the move (lines 11 to 15).
      2.1.3. Simulated annealing allows randomly accepting bad moves, in order to avoid local minimum results (line 13).

The pseudo-code of the general routine is presented in Figure 1.

```
1.  DSMClustering(matrix)
2.
3.  modules = createModules(matrix)
4.  tCCost = calcCoordCost(modules)
5.  counter = 0
6.  while isImproving()
7.    for i = 1 to size(DSM) * times
8.        element = pickRandomElement(matrix)
9.        bestModule = bid(modules, element)
10.       Move(element, bestModule)
11.       newCost = calcCoordCost(modules)
12.       if newCost <= tCCost or RandAccept()
13.           tCC = newCost
14.       else UndoMove()
```

**Figure 1. Original DSMC algorithm**

Figures 2 and 3 respectively show the pseudo-code for the calculations of *bid* and *TCC*.

```
1.  bid(modules, elem)
2.
3.  bestModule = NIL
4.  bestBid = -1;
5.  for each module mod ∈ modules
6.      bid = 0
7.      for each Member memb ∈ Members(mod)
8.          bid += DSM_{elem, memb}
9.      bid = bid^{powdep} /size(mod)^{powbid}
10.     if bid > bestBid
11.         bestBid = bid
12.         bestModule = mod
13. return bestModule
```

**Figure 2. Best *bid* calculation**

```
1.  calcCoordCost(modules)
2.
3.  for each member mb1 ∈ DSM
4.      for each member mb2 ∈ DSM
5.          // get element parent module
6.          Module mod1 = getModule(mb1)
7.          Module mod2 = getModule(mb2)
8.          cost = DSM_{mb1,mb2} + DSM_{mb2,mb1}
9.          if mod1 = mod2
10.             intraCost += cost*Size(mod1)^{powcc}
11.         else
12.             extraCost += cost*Size(DSM)^{powcc}
13. totalCost = intraCost + extraCost
14. return totalCost
```

**Figure 3. *TCC* calculation**

## 3. Research design

Our research methodology consisted of the following activities: (1) first, we implemented the original DSMC algorithm as described in section 2; (2) then we analyzed the algorithm performance, which led us to propose an optimization to improve its performance; (3) later, we tuned the algorithm parameters in order to better fit them to software clustering purposes; (4) and, finally, we conducted an empirical case study where we compared DSMC to other clustering algorithms.

### 3.1. Algorithm implementation

The DSMC algorithm was implemented in our Design Suite toolset, inside the Design Abstractor tool. This tool reads a design from a GXL graph file [9] and stores it in memory as a graph data structure. A design graph is converted to a DSM for executing the algorithm as explained in section 2.

### 3.2. Performance analysis and algorithm improvement

When we started our initial experiments with DSMC, we faced a scalability issue. For instance, in a design with around 1000 design entities, DSMC took around one week to converge to good clusterings. Although the complexity of the algorithm is polynomial, with much larger designs it may become unfeasible. Thus, additional work to optimize the calculations involved in DSMC might prove relevant for larger software. Analytical performance analysis can be useful to realize where the bottlenecks are.

Given a DSM with size $n$, and a set of output modules C, we determined the complexity of the algorithm as follows:
1. In the **calcCoordCost** routine, we can assume that finding the module the element belongs to and the access to the DSM cells is possible in constant time $O(1)$. Therefore, the routine has complexity $O(n^2)$, derived from the two nested loops.
2. For the **bid** routine, initially |C| is equals to $n$. So, we can define the its complexity as $O(n \cdot |C|)$, or simply $O(n^2)$.
3. The **DSMClustering** routine first calls the **createModules** subroutine, which takes $O(n)$ for creating one cluster for each member. After that, it reaches two nested loops. For the sake of simplicity, we may assume that the outer *while* loop runs in constant time $O(1)$. Taking into account the **for** loop and subroutine calls, this routine is $O(n + n \cdot (n^2 + n^2)) = O(n + n \cdot 2n^2) = O(n + n^3) = O(n^3)$.

We noticed that we could optimize the **calcCoordCost** routine by removing one of its loops.

In the **DSMClusterer** routine, *TCC* was being fully recalculated for each entity move. This calculation is not needed, since we simply want to **update** the old value of *TCC*.

Suppose an entity *mb* moves from module $m_1$ to module $m_2$. To correct the *TCC* value, it is enough to partially update the *ICC* and the *ECC*. It just takes 4 steps: 1) subtracting the intra cluster cost that *mb* caused in $m_1$, 2) adding the intra cluster cost that *mb* will cause within $m_2$, 3) adding an extra cluster cost now caused by *mb* in $m_1$ and 4) subtracting the extra cluster cost that *mb* ceased to cause in $m_2$.

Figures 4 and 5 present the new **DSMC** routine and its subroutine **updateTCC**.

```
15. DSMC(matrix)
16.
17. modules = createModules(matrix)
18. tCCost = calcCoordCost(modules)
19. counter = 0
20. while (IsImproving())
21.   for i = 1 to size(DSM) * times
22.     elem = ChooseRandomElement(matriz)
23.     old = GetModule(element)
24.     new = bid(modules, element)
25.     Move(elem, old, new)
26.     newCost = UpdateTCC(elem, old, new)
27.     if newCost <= tCC or RandAccept()
28.       totalCoordCost = newCost
29.     else UndoMove()
```

**Figure 4. Adapted DSMC algorithm**

```
1.  updateTCC(elem, old, new)
2.
3.  for each member mb1 ∈ old
4.    oldInner = DSM_{elem,mb1} + DSM_{mb1,elem}
5.    intraCost -= oldInner*Size(old)^{powcc}
6.    extraCost += oldInner*Size(DSM)^{powcc}
7.  for each member mb2 ∈ new
8.    oldOuter = DSM_{elem,mb2} + DSM_{mb2,elem}
9.    intraCost += oldInner*Size(old)^{powcc}
10.   extraCost -= oldInner*Size(DSM)^{powcc}
11. totalCost = intraCost + extraCost
12. return totalCost
```

**Figure 5. *TCC* update calculation**

### 3.3. Parameter tuning

The DSMC algorithm is fully parameterized, either for calculating *TCC* or *bid*, or for running the main loop. Parameter tuning may help in adapting the algorithm to a desired context and in better understanding and controlling its partially random behavior.

To improve *TCC* calculations, parameter *powcc* has to be tuned. The higher it is, the more larger clusters are penalized. On the other hand, *bid* calculations are parameterized by parameters *powdep* and *powbid*. The former may be used to emphasize, when attracting a design entity to a cluster, the power of their interactions. The latter serves for penalizing bids from larger clusters. Finally, three parameters are of interest for tuning the main loop: *times*, *randAccept* and *convergenceThreshold*. The first stands for the average number of iterations each entity is moved to a different cluster. Since entities are randomly chosen to be moved, repeating the loop for the DSM size allows, on average, that each entity is moved once. Nonetheless, some moves may not be the best due to the stochastic nature of the algorithm. Multiplying the DSM size by the parameter *times* allows more freedom of better choosing the most adequate clusters for each entity. The second parameter, *randAccept*, allows avoiding local minimum results through simulated annealing, by means of accepting bad moves once in each randAccept times. And the third parameter, *convergenceThreshold*, may be used to regulate the search process until no significative changes are made.

In order to find the best values for the above parameters, we performed a tuning task by changing their values in a predefined range and applying each of these values for fifty consecutive runs over five open source projects. We simplified the process, treating each parameter as independent from the others. Default values originally proposed by Fernandez [5] were used for the remaining parameters while tuning the first one. For the following parameters, the process was repeated except for changing the just tuned parameter to its new value.

Outputs were compared to a set of previously defined criteria, namely, authoritativeness and non-extremity [25], to decide the best values for the parameters. In summary, parameters that produced the most similar clusterings to authoritative decompositions (authoritativeness) and the least extreme clusterings (non-extremity, i.e., neither too large nor too small clusters) were chosen.

### 3.4. Case study

A case study was accomplished with software systems available from SourceForge, an open source software repository [21]. All studied systems were developed in Java. Their *.jar* files were input to design extraction after removing third-party libraries.

Latest versions of fifteen stable systems were input to the experiment. The research question was: how different clustering algorithms behave in terms of authoritativeness and extremity for a variety of different stable systems.

Table 1 shows a summary of characteristics from these systems.

**Table 1. Latest versions of stable systems**

| # | System | Size (KLOC) | Level | # of Nodes | # of Edges | Graph Size (KB) |
|---|---|---|---|---|---|---|
| 1 | JUnit 4.5 – compact | 2.3 | Extract | 614 | 1409 | 420 |
|   |   |   | Design | 23 | 70 | 21 |
| 2 | VilloNanny 1.0.0 | 2.9 | Extract | 435 | 1218 | 319 |
|   |   |   | Design | 25 | 79 | 23 |
| 3 | EasyMock 2.4 | 5.8 | Extract | 831 | 2128 | 590 |
|   |   |   | Design | 63 | 192 | 55 |
| 4 | PDF SaM 1.0.1 | 8.9 | Extract | 1144 | 2851 | 817 |
|   |   |   | Design | 68 | 176 | 56 |
| 5 | PJirc 2.2.1 | 28.5 | Extract | 2260 | 6966 | 1708 |
|   |   |   | Design | 133 | 419 | 118 |
| 6 | SweetHome 3D 1.3.1 | 38.1 | Extract | 5340 | 17039 | 4417 |
|   |   |   | Design | 97 | 518 | 127 |
| 7 | Jvlt 1.1.1 | 20.7 | Extract | 4224 | 13126 | 3298 |
|   |   |   | Design | 235 | 1279 | 304 |
| 8 | JEdit 4.2 | 140.7 | Extract | 7931 | 26938 | 6580 |
|   |   |   | Design | 234 | 1496 | 343 |
| 9 | Robocode 1.6.0.1 | 53.6 | Extract | 7563 | 21883 | 5627 |
|   |   |   | Design | 250 | 1184 | 291 |
| 10 | Jgnash 1.11.7 | N/A | Extract | 9754 | 33105 | 7949 |
|   |   |   | Design | 319 | 1905 | 442 |
| 11 | JabRef 2.4b2 | 109.2 | Extract | 11259 | 33990 | 8703 |
|   |   |   | Design | 461 | 2598 | 614 |
| 12 | JfreeChart 1.0.10 | 289.9 | Extract | 38301 | 129289 | 31805 |
|   |   |   | Design | 546 | 2790 | 668 |
| 13 | JavaGroups 2.6.3.GA | 124.9 | Extract | 22286 | 70097 | 17354 |
|   |   |   | Design | 554 | 3059 | 718 |
| 14 | PMD 4.2.3 | 80.7 | Extract | 42923 | 123907 | 32749 |
|   |   |   | Design | 569 | 3189 | 749 |
| 15 | FindBugs 1.3.5 | 170.3 | Extract | 54164 | 152377 | 40997 |
|   |   |   | Design | 967 | 6180 | 1419 |

Each of the four clustering algorithms is shortly described below:
- edge betweenness clustering (*eb*): clusters graphs based on edge betweenness. The betweenness of an edge is the extent to which that edge lies along shortest paths between all pairs of nodes. Edges which are least central to clusters are progressively removed until the clusters are separated [6];
- k-means clustering (*km*): clusters entities into a specified number of clusters, based on their proximity (in our case, the Jaccard distance) in d-dimensional space, using the k-means algorithm [7];
- modularization quality clustering (*mq*): finds clusters through optimization of a modularization quality function that maximizes cohesion and minimizes coupling [15];
- design structure matrix clustering (*dsm*): the algorithm thoroughly described in this work.

### 3.4.1. Evaluation criteria

A *software clustering* may be formally defined as the partitioning of a set of design-level entities. *Similarity* between partitions expresses how close they are to each other. Given two partitions $A$ and $B$, $MoJo(A, B)$ is the number of entity moves plus the number of cluster joins needed to transform $A$ into $B$ [25]. In order to measure similarity, one derived relative quality measure could be:

$$MoJoSim(A,B) = 1 - \frac{MoJo(A,B)}{n}$$

where $n$ is the number of entities to be clustered.

Wu and Holt have previously derived two criteria to evaluate aggregations which are used in this work: authoritativeness and extremity [25].

#### 3.4.1.1. Authoritativeness

One important measure of a software clustering algorithm utility is how close its resulting partition resembles one logical view created by an expert. To compare a partition $P$ generated by a clustering algorithm to an authoritative partition $P_A$, we measure $MoJoSim(P,P_A)$. We use available development views (Java package decompositions) as our expert decompositions.

#### 3.4.1.2. Non-Extremity of cluster distribution

A desirable property of an architectural clustering is that a cluster should resemble architectural components. Neither huge clusters nor singletons are usual in architectural components. Wu et al. [25] proposed a measure called non-extreme distribution (*NED*), defined as:

$$NED = \frac{\sum_{i=1, i \text{ not extreme}}^{k} n_i}{n}$$

where $k$ is the number of clusters in the partition, $n_i$ is the size of cluster $i$ and $n$ is the total of entities to be clustered. We used 5 and 20 as the lower and upper limits, respectively, for non-extreme clusters.

## 4. Results

The results derived from the tuning process and from the complexity analysis over the new DSMC algorithm are shown below. Also, the comparison with other algorithms is present.

### 4.1. Performance analysis for DSMC

Analyzing the adapted version of DSMC presented in Figures 4 and 5, it is easy to see that the **updateTCC** routine is $O(n)$, which runs faster than **calcCoordCost** in the original version. Actually, this improvement does not affect the worst case run-time for the **DSMC** routine, which is still $O(n^3)$ due to the **bid** subroutine. On the other hand, considering the systems we studied, the **bid** routine has a much lower expected time than its worst case, since $|C|$ decreases to a value much lower than $n$. The actual decrease is unknown, but for the typical cases we studied, there is a strong reduction in the constants multiplying the higher order term. Hence, the adapted **DSMC** is still $O(n^3)$, but the constant $c$ multiplying the $n^3$ term is much lower than in the original algorithm.

As an example of this improvement, table 3 summarizes data contained in two execution logs, one for the original algorithm and another for the adapted one. Log data include *TCC* values, number of generated clusters, total number of iterations, mean iteration time, and total run-time. All shown measures are average values over thirty algorithm runs until convergence was reached.

**Table 2. Run time log data for both DSMC algorithm variants**

| Log Data | Original | Adapted |
|---|---|---|
| TCC | 1700738.0 | 1500287.0 |
| Nº of clusters | 39 | 28 |
| Iterations count | 224 | 448 |
| Iteration mean time | 723.0ms | 5.0ms |
| Total run time | 162.0s | 2.3s |

### 4.2. Parameter tuning

Each parameter was varied in a range of probe values. For each of the five systems and for each parameter value, the algorithm was run fifty times and the average values of authoritativeness and non-extremity were calculated.

Figures 6 and Figure 7 respectively present *MojoSim* and *NED* average values for *powcc*, calculated as explained in 3.4.1. For visual clarity, values are plotted only for three systems.

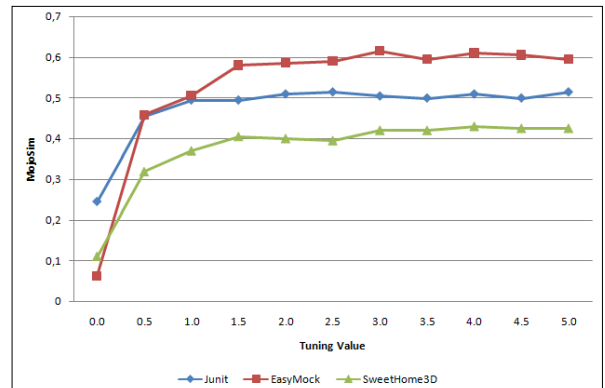

**Figure 6. Tuning *powcc* for authoritativeness**

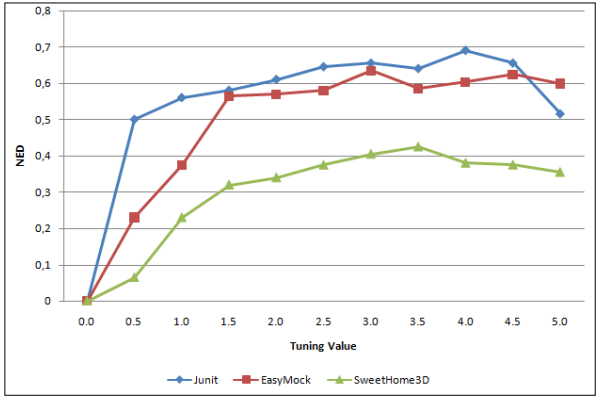

**Figure 7. Tuning *powcc* for non-extremity**

The chosen value for *powcc* was 3. A summary of our parameter tuning and a list of our chosen values is shown in table 3. A comparison of the chosen parameter values with the values used by Fernandez and Thebeau is shown in table 4.

**Table 3. Parameter tuning**

| Parameter | Range | Increment | Best Value |
|---|---|---|---|
| *powbid* | 0.0 – 5.0 | 0.5 | 1 |
| *powdep* | 1.0 – 9.0 | 2.0 | 5 |
| *powcc* | 0.0 – 5.0 | 0.5 | 3 |
| *times* | 1.0 – 10.0 | 1.0 | 4 |
| *randAccept* | 5.0 – 50.0 | 5.0 | 5 |

**Table 4. Parameter tuning comparison**

| Parameter | Fernandez | Thebeau | Ours |
|---|---|---|---|
| *powbid* | 1 | 1 | 1 |
| *powdep* | 0 | 1 | 5 |
| *powcc* | 1 | 4 | 3 |
| *times* | 2 | 2 | 4 |
| *randAccept* | 30 | 122 | 5 |

## 4.3. Case study

Results for the case study are shown below. Charts were plotted for authoritativeness and extremity. Relative measures were derived to better position each algorithm relatively to the others.

In a previous paper we compared these four algorithms in a different case study, where consecutive versions of four different open-source systems were submitted as input to these algorithms and results were compared in terms of authoritativeness, non-extremity and stability [2].

### 4.3.1. Relative measures

In order to compare data series, Wu et al. defined ordinal measures to rank two or more data series [25].

For two series $DS_i$ and $DS_j$, the relative measure *Above* is defined as:

$$Above(DS_i, DS_j) = \frac{|\{n | DS_i[n] > DS_j[n], 1 \le n \le |DS_i|\}|}{|DS_i|}$$

For *k* data series, the relative measure for a particular series $DS_i$ in relation to all the other *k* series is defined as:

$$Above(DS_i) = \sum_{j=1}^{k} Above(DS_i, DS_j)$$

### 4.3.2. Non-Extremity of Cluster Distribution

We calculated the non-extreme distribution (*NED*) measure for the case study, as defined in 3.4.1. Figure 9 shows the *NED* data series for each algorithm for the stable systems analysis.

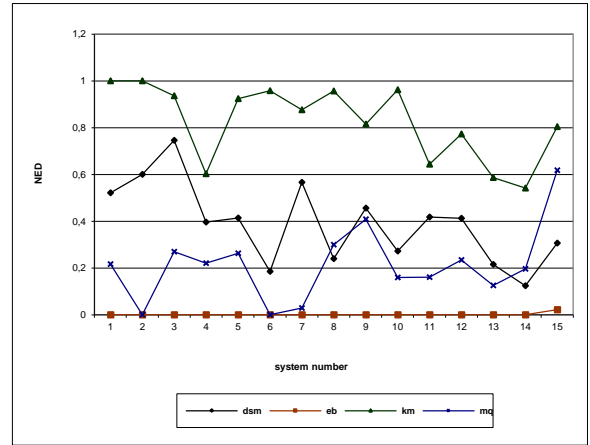

**Figure 9. *NED* scores for case study**

Table 5 shows the relative non-extremity measure *Above*.

The results suggest that *km* clustering performs best in terms of non-extremity, followed by *dsm* and *mq*, and *eb* performs worst. The number of clusters in *km* is parameterized to 10% of the number of entities, thus, easily forming non-extreme clusters. Furthermore, most clusterers produce low *NED* values (below 0.5), except for *km*, that produces, on average, medium *NED* values. Looking closely at the clusters formed, one can see that *dsm* and *mq* form some non-extreme clusters and many small clusters, while *eb* usually forms one huge cluster and many singletons.

**Table 5. Relative non-extremity scores**

| Algorithm | Above score |
|---|---|
| *eb* | 0.00 |
| *km* | 3.00 |
| *mq* | 1.07 |
| *dsm* | 1.80 |

### 4.3.3. Authoritativeness

For each system in the case study, we calculated the similarity between the partition *P* formed by the four studied algorithms and the authoritative partition $P_A$ formed by the package decomposition, using the *MoJoSim(P,$P_A$)* measure, as defined in 3.4.1. Figure 10 shows the *MoJoSim* data series for each algorithm for the stable systems analysis.

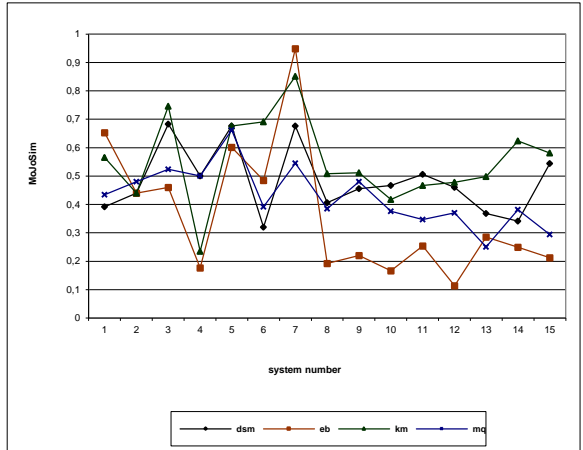

**Figure 10. *MoJoSim* authoritativeness scores for case study**

Table 6 shows the relative authoritativeness measure *Above*.

**Table 6. Relative authoritativeness scores**

| Algorithm | Above score |
|---|---|
| eb | 0.60 |
| km | 2.33 |
| mq | 1.20 |
| dsm | 1.53 |

The results point that *km* and *dsm* compete for best authoritativeness, with some advantage for *km*, being followed by *mq*. Algorithm *eb* ranks worst.

## 5. Discussion

We shall now proceed to a short discussion of our findings, in the light of our research design.

First, regarding performance analysis, we could notice that the original formulation of the algorithm, although of polynomial complexity, would take an unfeasible time to reach a good clustering for systems with more than, for instance, 1000 design entities (e.g.: classes, files). Since these are medium-sized systems, the DSMC algorithm would not be worth for large systems, which are usually the target of automatic clustering algorithms for architecture recovery. After making a slight change in the algorithm, we could make it work with large systems. For instance, in FindBugs, a system with 967 design entities, run time improved from more than one week to nearly two hours to converge. Anyway, the same issue arises with other algorithms that solve the clustering problem through optimization of an objective function such as the modularization quality clustering algorithm [15].

Parameter tuning was an important issue in improving DSMC to the architecture recovery domain. Previous work with DSMC was restricted to the domains of task scheduling and team interaction and the tuning performed was based on criteria for these domains. The use of the criteria of authoritativeness and non-extremity was a means of taking into account software architecture issues for parameter tuning. Since the DSMC offers a number of degrees of freedom through its parameters, using heuristics typical from software architecture should be a natural way of adjusting the algorithm to software architects' needs.

In summary, improvements in the algorithm consisted of an optimization to incrementally calculate the total coordination cost and the choice of parameter values to better fit clustering results to software architecture recovery criteria.

Obviously, there is a limitation on those quantitative criteria. The first one, authoritativeness, is limited by the availability of authoritative decompositions. In our case, package decompositions were the only available authority decision that we could hold on and they are limited by software designers' worries about reflecting the software logical structure in its packages. The second criterion, non-extremity, has two important drawbacks: first, non-extreme clusters are seen as the only good decomposition, although there are situations where small extreme clusters may be welcome, and, second, the lower and upper limits that separate good from bad clusters are empirical and not thoroughly validated.

Results for the case study drive, on the other hand, some interesting discussion on the quality and adequacy of DSMC and the other compared algorithms for tasks of architecture recovery. Quantitative results favor k-means, not only for authoritativeness but also for non-extremity. But an important issue in architecture recovery is that one does not usually know in advance the number of clusters to be recovered and that is a strong drawback of k-means, since it requires this value as a parameter. Both modularization quality (MQ) and DSMC take into account software engineering heuristics information hiding, high cohesion and low coupling in their objective functions, but with subtle differences. MQ is explicit in defining structural cohesion and coupling and tries to maximize the first and minimize the second. DSMC, on the other hand, treats both these issues in the *TCC* and *bid* functions. With more degrees of freedom through parameters in these functions, DSMC makes better figures than MQ.

Finally, another important issue that we will discuss only superficially here is about a more subjective quality of the clustering results. From the heuristics of MQ and DSMC, the resulting clusters are usually more cohesive and less coupled than in k-means. Since this is an important issue in software engineering, this theme deserves a more thorough investigation. Usually, automatic clustering results are a starting point for a software architect recovering a module view. Whether these more cohesive and less coupled clusters are a better starting point than the smaller number of not so cohesive clusters obtained by k-means is an issue to be solved by other quantitative criteria or through subjective qualitative studies.

## 6. Related Work

Software architecture recovery is a long-standing research subject, since the early nineties. A recent survey recovers prolific work in this area [18]. Another older study surveys the area of software architecture with focus on combating architectural degeneration, giving emphasis to architecture and design recovery [8].

Automatic clustering techniques aim to recover high-level abstractions with information available in low-level models and very few or no intervention from a reverse engineer. Models represented as graphs can be abstracted through graph clustering techniques. Global and local graph clustering methods are surveyed in [20]. Hartigan proposed the k-means algorithm to cluster vector data [7], which was later adapted to graph clustering. Girvan and Newman proposed a graph clustering method to find communities in social networks through the removal of edges with high betweenness [6]. Anquetil and Lethbridge used agglomerative hierarchical algorithms to automatically cluster software [1]. Maqbool and Babri proposed a weighted combined linkage algorithm to measure the distance between clusters and aggregate the most similar in an agglomerative fashion [16]. Mitchell and Mancoridis built a software clustering tool named Bunch that hosts a suite of graph clustering algorithms based on the optimization of a modularization quality function that rewards modules with high cohesion and low coupling [15, 17].

Fernandez represented design modules as a design structure matrix (DSM) and proposed an optimization algorithm to partition the design into clusters in order to minimize the coordination cost between teams that worked on the modules [5]. Sangal et al. applied DSMs to analyze software through the use of the Lattix LDM tool [19], which nowadays also partitions software design [12].

Other automatic recovery techniques not focused on clustering are based on concept analysis [24] and graph dominance [13].

## 7. Conclusion

This work presented an algorithm based on design structure matrices (DSMC) and applied it to the domain of software architecture recovery. The algorithm was based on previous prolific work on design structure matrices and was adapted to the area of software architecture recovery by means of performance analysis and parameter tuning. Through performance analysis, we realized a limitation of the optimization-based algorithm and adapted it to an incremental form in order to speed up convergence. Parameter tuning used quantitative criteria derived from software engineering heuristics in order to find clusters more typical of software architecture recovery tasks. A case study was performed to compare DSMC with other clustering algorithms. It outperformed edge betweenness clustering, an algorithm used in social networks, and also the modularization quality clustering, another optimization-based clustering algorithm. On the other hand, it was outperformed by the k-means clustering algorithm, although this algorithm requires in advance the knowledge of the number of clusters, which is not the case for DSMC.

Further work shall be pursued in terms of algorithm comparison using other quantitative criteria and also using qualitative feedback from software architects as to whether DSMC clustering and the other algorithms recover modules that make sense to software architects. In addition, DSMC variants with other heuristics than the ones used here are another field of research that might bring interesting results to software architecture recovery.

This work shows the capabilities of design structure matrices for abstraction activities. Not only are DSMs useful as a visualization technique or as a way to better organize or schedule product production, but they serve also as a powerful mathematical tool for abstracting software architecture modules, with comparable results with other approaches.

## References

[1] N. Anquetil and T. Lethbridge. Experiments with Clustering as a Software Remodularization Method. In *Proceedings of the Sixth Working Conference on Reverse Engineering*, pages 235–255, October 1999.

[2] R. A. Bittencourt and D. D. S. Guerrero. Comparison of Graph Clustering Algorithms for Recovering Software Architecture Module Views. In *Proceedings of the 13th European Conference on Software Maintenance and Reengineering*, 2009.

[3] T. R. Browning. Applying the design structure matrix to system decomposition and integration problems: A review and new directions. *IEEE Transactions on Engineering Management*, 48(3):292–306, August 2001.

[4] DSM Web. Website, 2009. http://www.dsmweb.org.

[5] C. I. G. Fernandez. *Integration Analysis of Product Architecture to Support Effective Team Co-location*. Master's Thesis (ME), MIT, 1998.

[6] M. Girvan and M. E. J. Newman. Community structure in social and biological networks. *Proceedings of the National Academy of Sciences*, 99(12):7821–7826, June 2002.

[7] J. A. Hartigan and M. A. Wong. A K-Means Clustering Algorithm. *Applied Statistics*, 28(1), 1979.

[8] L. Hochstein and M. Lindvall. Combating architectural degeneration: a survey. *Information and Software Technology*, 47(10):643–656, July 2005.

[9] R. C. Holt, A. Schürr, S. E. Sim, and A. Winter. GXL: A graph-based standard exchange format for reengineering. *Science of Computer Programming*, 60(2):149–170, 2006.

[10] S. Huynh, Y. Cai, Y. Song and K. Sullivan. Automatic modularity conformance checking. In *Proceedings of the 30th international Conference on Software Engineering*, 2008.

[11] R. Kazman, L. O'Brien and C. Verhoef. *Architecture Reconstruction Guidelines*. Technical Report CMU-SEI-2001-TR-026, 2001.

[12] Lattix, Inc. Website, 2008. http://www.lattix.com/

[13] J. Lundberg and W. Löwe. Architecture recovery by semi-automatic component identification. *Electronic Notes in Theoretical Computer Science*, 82(5):98–114, 2003.

[14] A. MacCormack, J. Rusnak and C. Y. Baldwin. Exploring the Structure of Complex Software Designs: An Empirical Study of Open Source and Proprietary Code. Management Science. 52(7):1015-1030, July 2006.

[15] S. Mancoridis, B. S. Mitchell, C. Rorres, Y. Chen, and E. Gansner. Using automatic clustering to produce high-level system organizations of source code. In *Proc. 6th International Workshop on Program Comprehension*, 1998.

[16] O. Maqbool and H. A. Babri. The weighted combined algorithm: A linkage algorithm for software clustering. In *Proceedings of the 8th European Conference on Software Maintenance and Reengineering*, 2004.

[17] B. S. Mitchell and S. Mancoridis. On the automatic modularization of software systems using the bunch tool. *IEEE Transactions on Software Engineering*, 32(3):193–208, March 2006.

[18] D. Pollet, S. Ducasse, L. Poyet, I. Alloui, S. Cimpan, and H. Verjus. Towards A Process-Oriented Software Architecture Reconstruction Taxonomy. *In Proceedings of the 11th European Conference on Software Maintenance and Reengineering*, 2007.

[19] N. Sangal, E. Jordan, V. Sinha, and D. Jackson. Using dependency models to manage complex software architecture. In *Proc. 20th Annual ACM SIGPLAN Conference on Object Oriented Programming Systems, Languages and Applications*, pages 167–176, October 2005.

[20] S. E. Schaeffer. Graph Clustering. *Computer Science Review*, 1(1):27-64, August 2007.

[21] SourceForge. Website, 2008. http://sourceforge.net.

[22] K. Sullivan, Y. Cai, B. Hallen and W. Griswold. The Structure and Value of Modularity in Software Design. In *Proc. 8th European Software Engineering Conference / 9th ACM SIGSOFT International Symposium on Foundations of Software Engineering*, 2001.

[23] C. I. G. Thebeau. *Knowledge Management of System Interfaces and Interactions for Product Development Processes*. Master's Thesis (ME), MIT, 2001.

[24] T. Tilley, R. Cole, P. Becker, and P. Eklund. A Survey of Formal Concept Analysis Support for Software Engineering Activities. In *Proc. 1st International Conference on Formal Concept Analysis*. Springer, 2003.

[25] J. Wu, A. E. Hassan, and R. C. Holt. Comparison of Clustering Algorithms in the Context of Software Evolution. In *Proc. 21st IEEE International Conference on Software Maintenance*, 2005.